\documentclass[aps,preprint,amsmath,amssymb]{revtex4}
\usepackage{psfig}
\usepackage{graphicx}

\begin{document}

\title{\bf
What can we learn from $B^{+}\to D^{(*)-}_{s} K^{+} \pi^{+}$,
$B_{d}\to D_{s}^{(*)-} K^{0} \pi^{+}$ and $B_{d}\to D_{s}^{(*)-}
K^{*+}$ decays?}
\date{\today}
\author{  \bf  Chuan-Hung Chen\footnote{Email: chchen@phys.sinica.edu.tw} }

\vskip1.0cm

\address{  Institute of Physics, Academia Sinica,
Taipei, Taiwan 115, Republic of China}

\begin{abstract}
We study the nonresonant three-body decays of $B^{+}\to
D^{(*)-}_{s} K^{+} \pi^{+}$ and $B_{d}\to D_{s}^{(*)-} K^{0}
\pi^{+}$.
We find that these decays
can provide the information on the time-like form factors
of $D^{(*)}_{s} K$.
We also explicitly investigate $B_{d}\to D_{s}^{(*)-} K^{*+}$ decays
by discriminating the nonresonant contributions
with the unknown $D^{(*)}_{s}$ wave functions being fixed by the
measured mode of $B_{d}\to D_{s}^{-} K^{+}$.
\end{abstract}
\maketitle

Three-body decays of $B$ meson  have recently been noticed in the
experiments by the analyses of the Dalitz plots and
invariant mass distributions \cite{Belle-PRD}. The study of
the three-body decays provides not only the method to extract the CP
violating phase angles \cite{CP}, but also the way to understand
or search the uncertain particle states, such as
$f_{0}(400-1200)$, $f_{0}(980)$, $a_{0}(980)$ \cite{Chen-PRD} and
glueballs \cite{CHT}. Moreover, it also helps us to build up
the QCD approach for the nonresonant three-body decays
\cite{ChenLi}. As known that the charmless three-body decays of
$B$ meson are dominated by the so-called quasi-two-body decays
\cite{CY}, it is not  easy to discriminate the nonresonant
states from resonant ones. However, this may not be the case in
those final states with charmed mesons.

It has been demonstrated by Belle \cite{Belle-PLB} that $B_{d} \to
D^{(*)-} \bar{K}^{0} K^{+}$ decays are actually  dominated by the
three-body modes because the $\bar{K}^{0} K^{+}$ system is
confirmed to be an $J^{P}=1^{-}$ state
by the analysis of angular dependence.
The production of the decays can be thought easily by the
consequence of $b\to c \bar{u} d$, while $\bar{K}^{0} K^{+}$ is
produced by the created $s\bar{s}$ pair and $\bar{u} d$.
Therefore, if annihilation topologies are neglected, the dominant
topologies for the decays correspond to $B_{d}\to D^{(*)-}$ and a
outgoing pair of $\bar{K}^{0} K^{+}$. The formers are described by
$B_{d}\to D^{(*)-}$ form factors, calculated by some QCD
approaches such as perturbative QCD (PQCD), quark model, QCD sum
rules and light-cone QCD sum rules, while the latters respond the
times-like form factors which can be fixed via the connection to
electromagnetic form factors and fitting with experiments
\cite{Balakin}. Although time-like form factors of $KK$, denoted
by $F^{0\to KK}$, are not easy to be formulated in theory,
$B_{d}\to D^{(*)-}\bar{K}^{0} K^{+}$ is a good candidate to study
the nonresonant three-body decays. According to the observations
of Belle, we know that the branching ratios (BRs) of $B_{d} \to
D^{(*)-} \bar{K}^{0} K^{+}$ decays are of ${\cal O}(10^{-4})$. It
means that the effects of $F^{0\to KK}$ are not small. Moreover,
following the analysis of Ref. \cite{CHST}, we see that the peaks
in the $KK$ spectra of the decays locate at around $1.5$ GeV. This
can be understood that the region is actually  governed by PQCD
where the proper hard scale is around $\sqrt{\bar{\Lambda} m_{B}}$
\cite{ChenLi} with $\bar{\Lambda}=m_{B}-m_b$.
By taking $\bar{\Lambda}\approx 0.48$ and $m_{B}=5.28$ GeV, the
value of $\sqrt{\bar{\Lambda} m_{B}}$ is $1.6$ GeV, quite close to
the consequence of Ref. \cite{CHST}. In some senses, PQCD approach
can deal with the three-body decays by combing with the
experimental fittings of time-like form factors.

Inspired by the large BRs of  $B_{d}\to D^{(*)-}\bar{K}^{0} K^{+}$
decays, one can speculate that  the three-body decay
related to form factors $\langle DP|V_{\mu}-A_{\mu}|B\rangle$
can be also large, saying
${\cal O}(10^{-5}-10^{-4})$,
where $D(P)$ is the charmed (pseudoscalar) meson and $V_{\mu}(A_{\mu})$
is the vector (axial-vector) current. In theoretical viewpoints, the
question is hard to answer since there are too much unknown form factors
involved and no direct experimental data related to them. However,
we still can give some conjectures on the relevant decays.
Firstly, in terms of the concept of two-meson wave functions
\cite{MP}, the $DP$ system could be described by a set of wave
functions for $\langle DP|V_{\mu}-A_{\mu}|B\rangle$ and they can
be related to the time-like form factors of $DP$, denoted by
$F^{0\to DP}$. Therefore, if the $K$ meson is massless
particle, the threshold invariant mass, expressed by $\omega$, to
generate the $KK$ pair is about $0$. However, unlike
the case of the $KK$ pair production, since charmed meson is a massive
particle, to produce the $DP$ pair it should
start from $m_{D}$. By assuming that the peak of the
$DP$ pair spectrum is around $m_{D}+\sqrt{\bar{\Lambda} m_{B}}$,
 we can expect
that the BR associated with $F^{0\to DP}$ form factors should be
smaller than that associated with $F^{0\to KK}$ because the
dominant form factors have been shifted to a larger $\omega$
region and their values are small, compared to $F^{0\to KK}$ at
$\sqrt{\bar{\Lambda} m_{B}}$.
Moreover, if the
third particle of the involving three-body decay is a light meson,
although the allowed $\omega$ of $DP$ could reach the value of
$m_{B}$ ( it is $m_{B}-m_D$ in the $B_{d}\to D^{(*)-} \bar{K}^{0}
K^{+}$ decays), due to the suppression of phase space factor
$(1-\omega^2/m^2_{B})$, the effects of the large $\omega$ are not
important. Therefore, the available phase space is smaller than
that in the decays of $B_{d}\to D^{(*)-} \bar{K}^{0} K^{+}$. Hence we
conjecture that the BR of $B\to DPP'$
should be smaller than
those of $B_{d}\to D^{(*)-} \bar{K}^{0} K^{+}$ decays,
where $DP$ system and
$B$ meson have the same light spectator. Note that
the chosen examples have the same weak Wilson coefficients (WCs).

\begin{figure}
\includegraphics*[width=3.6
  in]{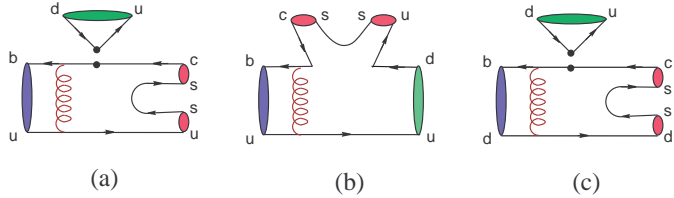} 
\caption{The topologies for the nonresonant three-body decays (a,
b) $B^{+}\to D^{(*)-}_{s} K^{+} \pi^{+}$ and (c) $B_{d}\to
D^{(*)-}_{s} K^{0} \pi^{+}$. The dots denote the weak vertices.}
\label{topology}
\end{figure}

Because there is no any direct information on $F^{0\to DP}$, in
order to confirm our conjectures, we suggest that the observations
of $B^{+}\to D^{(*)-}_{s} K^+ \pi^{+}$ and $B_{d}\to D^{(*)-}_{s}
K^0 \pi ^+$, illustrated by Fig. \ref{topology}, can help us to
find the answer. Since the former modes correspond to pure
three-body decays and once they are measured in
experiments, we immediately know what the effects of $F^{0\to DP}$
are. However, besides the nonresonant three-body decays, the
decays of $B_{d}\to D^{(*)-}_{s} K^0 \pi ^+$ also involve resonant
states $B_{d}\to D^{(*)-}_{s} K^{*+}(K^{*+}\to K^{0} \pi^{+})$.
Recently, Belle \cite{Belle} and Babar \cite{Babar} have measured
the relevant two-body decay $B_{d}\to D^{-}_{s} K^{+}$ to be
$(4.6^{+1.2}_{-1.1}\pm1.3)\times 10^{-5}$ and $(3.2 \pm 1.0 \pm
1.0)\times 10^{-5}$, respectively. One expects that the $B_{d}\to
D^{(*)-}_{s} K^{*+}$ decays should have the same magnitudes in BR.
Probably, the suggested three-body decays also have the same BRs
in order of magnitudes. In order to understand more on the nonresonant
parts in $B_{d}\to D^{(*)-}_{s} K^0 \pi ^+$, it is
important to know how large the contributions are from
quasi-two-body decays. In this paper, we want to make
a detailed analysis on
$B_{d}\to D^{(*)-}_{s} K^{*+}$ decays. 

Since the considered decays correspond to the $b\to c \bar{d} u$
transition, we describe the effective Hamiltonian as
\begin{eqnarray}
H_{{\rm eff}}&=&\frac{G_{F}}{\sqrt{2}}\sum_{q=u,c}V_{q}\left[
C_{1}(\mu ){\cal O}_{1}^{(q)}+C_{2}(\mu ){\cal O}
_{2}^{(q)}\right]
\nonumber\\
{\cal O}_{1}^{(q)} &=& \bar{d}_{\alpha} q_{\beta}
\bar{c}_{\beta} b_{\alpha}\,,\
\ \   {\cal O}_{2}^{(q)} \;=\;
\bar{d}_{\alpha} q_{\alpha} \bar{c}_{\beta} b_{\beta}
\label{eff}
\end{eqnarray}
where $\bar{q}_{\alpha} q_{\beta}=\bar{q}_{\alpha} \gamma_{\mu}
(1-\gamma_{5}) q_{\beta}$, $\alpha(\beta)$ are the color indices,
$V_{q}=V_{qd}^{*}V_{cb}$ are the products of the CKM matrix
elements \cite{CKM}, and $C_{1,2}(\mu )$ are the WCs \cite{BBL}.
Conventionally, the effective WCs of $a_{2}=C_{1}+C_{2}/N_{c}$ and
$a_{1}=C_{2}+C_{1}/N_{c}$ with $N_{c}=3$ being color number are
more useful. It is known that the difficulty for studying
exclusive hadron decays is from the calculations of matrix
elements. In order to handle the hadronic effects, we employ the
PQCD approach in which the transition matrix element is described
by the convolution of hadron wave functions and the hard amplitude
of the valence quarks \cite{LB,Li}. Although the PQCD approach
suffers singularities from end-point region, they could be smeared
after the threshold and the $k_{T}$ resummation effects are
included. The latter arises from the introduction of the parton
transverse momentum \cite{KLS-PRD,CKL-PRD}. In the literature, the
applications of PQCD to exclusive $B$ meson decays, such as
$B\rightarrow K \pi$ \cite{KLS}, $B\rightarrow \pi \pi(KK) $
\cite{LUY,CL-PRD,Chen-PLB520}, $B\rightarrow \phi \pi(K)$
\cite{Melic,CKL-PRD}, $B\rightarrow \eta^{(')} K$ \cite{KS},
$B\rightarrow \rho K$ \cite{Chen-PLB525} decays and $B\to K^{(*)}
\ell^{+} \ell^{-}$ \cite{CQ-PRD}, have been studied and found that
all of them are consistent with the current experimental data
\cite{SU,Keum}. We think that the same approach can be also used
to the considered cases here.

It has been shown that by the reality of hierarchy
$m_{B}>>m_{D^{(*)}_{s}}>>\bar{\Lambda}$, the  distribution
amplitudes of $B(D^{(*)}_{s})$ mesons can be described by
\cite{Li-P}
\begin{eqnarray}
\langle 0|\bar{b}(0)_{j} d(z)_{l}|B,p_{1}\rangle
&=&\frac{1}{\sqrt{2N_{c}}}\int^{1}_{0} dx e^{-ixp_{1}\cdot z}
 \Big\{ [ \slash \hspace{-0.2cm}
p_{1} +m_{B} ]_{lj}\gamma_{5} \Phi_{B}(x)\Big\},
\end{eqnarray}
\begin{eqnarray}
\langle D_{s},p_{2}|\bar{d}(0)_{j} c(z)_{l}|0\rangle
&=&\frac{1}{\sqrt{2N_{c}}}\int^{1}_{0} dx e^{ixp_{2}\cdot z}
 \Big\{ \gamma_{5}[ \slash \hspace{-0.2cm}
p_{2} + m_{D_{s}} ]_{lj} \Phi_{D_{s}}(x)\Big\}, \nonumber \\
\langle D^{*}_{s},p_{2}|\bar{d}(0)_{j} c(z)_{l}|0 \rangle
&=&\frac{1}{\sqrt{2N_{c}}}\int^{1}_{0} dx e^{ixp_{2}\cdot z}
\Big\{ \slash \hspace{-0.18 cm} \varepsilon_{2}  [ \slash
\hspace{-0.2cm} p_{2} + m_{D^{*}_{s}} ]_{lj}
\Phi_{D^{*}_{s}}(x)\Big\},\label{dad}
\end{eqnarray}
where $\varepsilon_{2\mu}$ is the polarization vector of
$D^{*}_{s}$ meson, the normalizations of wave functions are taken
to be $\int^{1}_{0}
dx\Phi_{B(D^{(*)}_{s})}(x)=f_{B(D^{(*)}_{s})}/2\sqrt{2N_{c}}$ and
$f_{B(D^{(*)}_{s})}$ are the corresponding decay constants.
Although the decay constants and wave functions of $D^{*}_{s}$
 between longitudinal and transverse polarizations are
different generally, for simplicity, we assume that they are the
same. Since the effects of transverse polarization parts are
always related to the factor $r_{2}=m_{D^*_{s}}/M_{B}$, one
expects that their contributions are much smaller than those from
longitudinal parts. As to the distribution amplitude of the $K^*$
meson, we refer to the results derived from QCD sum rules
\cite{BBKT} and summarize them as
\begin{eqnarray}
\langle 0|\bar{u}(0)_{j}
s(z)_{l}|K^{*},p_{3}\rangle&=&\frac{1}{\sqrt{2N_c}}\int^{1}_{0} dx
e^{-ixp_{3}\cdot z} \Big[m_{K^*}[\slash \hspace{-0.14
cm}\varepsilon_{3L}]_{lj}\Phi_{K^*}(x) +[\slash \hspace{-0.14
cm}\varepsilon_{3L}\slash \hspace{-0.18 cm} p_{3}]_{lj}
\Phi_{K^*}^{t}(x)\nonumber\\&&  +m_{K^*} [I]_{lj}\Phi_{K^*}^s(x)+
m_{K^*}[\slash \hspace{-0.14 cm}
\varepsilon_{3T}]_{lj}\Phi_{K^*}^v(x)\nonumber \\&&+ [\slash
\hspace{-0.14 cm}\varepsilon_{3T} \slash \hspace{-0.18 cm}
p_{3}]_{lj} \Phi_{K^*}^T(x) +m_{K^*}[\slash \hspace{-0.18 cm}{\cal
C}]_{lj}\Phi_{K^*}^a(x)\Big], \label{ksf}
\end{eqnarray}
where $\varepsilon_{3L(T)}$ denote the longitudinal (transverse)
polarization vectors of $K^*$ meson, ${\cal C}_{\mu}=$ $i
\epsilon_{\mu\nu\rho\sigma}$ $\varepsilon_{3T}^\nu p_{3}^\rho
n_{-}^\sigma/{p_{3}\cdot n_{-}}$ in which $n_{-}$ is parallel to
the large component of $p_{3}$, and $\Phi_{K^*}^{(T)}$ correspond
to twist-2 wave functions while the remains stand for the twist-3
ones. In addition, in terms of light-cone coordinate, the momenta
of various mesons and the light valence quarks inside the
corresponding mesons are assigned as:
$p_{1}=m_{B}/\sqrt{2}(1,1,\vec{0}_{T})$,
$k_{1}=m_{B}/\sqrt{2}(x_{1},0,\vec{k}_{1T})$;
$p_{2}=m_{B}/\sqrt{2}(1,r^{2}_{2},\vec{0}_{T})$,
$k_{2}=m_{B}/\sqrt{2}(x_{2},0,\vec{k}_{2T})$;
$p_{3}=m_{B}/\sqrt{2}(0,1-r^{2}_{2},\vec{0}_{T})$,
$k_{3}=m_{B}/\sqrt{2}(0,(1-r^{2}_{2})x_{3},\vec{k}_{3T})$.

\begin{figure}[h]
\includegraphics*[width=2.2
in]{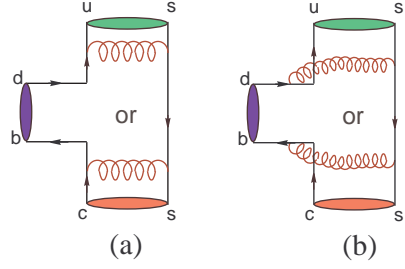} 
\caption{The topologies (a) factorizable (b) nonfactorizable
effects for the decays $B_{d}\to D^{(*)-}_{s} K^{*+}$.}
\label{two-body}
\end{figure}

As usual, we have the decay rate for $B_{d}\to D^{-}_{s}
K^{*+}$ decay as
\begin{eqnarray*}
\Gamma=\frac{G^{2}P_{c}m^{2}_{B}}{16\pi } |A_{D_{s}^{-}
K^{*+}}|^{2},
\end{eqnarray*}
where $P_c\equiv |p_{2z}|=|p_{3z}|$
and the amplitude of $A_{D_{s}^{-} K^{*+}}$ is
given by
\begin{eqnarray}
A_{D_{s}^{-} K^{*+}}=f_{B}F_{D_{s}^{-} K^{*+}}+M_{D_{s}^{-}
K^{*+}}.
\label{Eq5}
\end{eqnarray}
The first (second) term in Eq. (\ref{Eq5}) comes from the factorizable
(nonfactorizable) contributions, illustrated by Fig.
\ref{two-body}a (\ref{two-body}b).
The hard amplitudes
$F_{D_{s}^{-} K^{*+}}$ and $M_{D_{s}^{-} K^{*+}}$ are expressed by
\begin{eqnarray}
F_{D_{s}^{-} K^{*+}}&=&-8 \pi C_{F} m^{2}_{B} \int_{0}^{1} d[x]
\int_{0}^{\infty} [b]d[b] \Phi_{D_{s}}( x_{2})
\nonumber\\
&& \Big\{ \Big[(r^{2}_{2}-(1-2r^{2}_{2})x_{3})\Phi_{K^*}(\zeta)
+r_{2}r_{3}\Big((1-2x_{3})\Phi^{t}_{K^*}(\zeta)+(1+2x_{3})\Phi^{s}_{K^*}(\zeta)
\Big) \Big] \nonumber \\ && \times {\cal
E}_{a}^{1}(t^{1}_{a})+\Big[
(1-r^2_{2})x_{2}\Phi_{K^*}(\zeta)-r_{2}r_{3}(1+x_{2})\Phi^{s}_{K^*}(\zeta)\Big]{\cal
E}_{a}^{2}(t^{2}_{a}) \Big\}, \label{fa} \\
M_{D_{s}^{-} K^{*+}}&=&16\pi C_F m_{B}^{2}\sqrt{2N_{c}}
\int_{0}^{1} d[x] \int_{0}^{\infty} [b]d[b] \Phi_{B}(x_{1},b_{1})
\Phi_{D_{s}}( x_{2}) \nonumber \\
&& \Big\{ \Big[ -(r_{2}^{2}x_{2}+(1-2r_{2}^{2})x_{3})
 \Phi_{K^*}(\zeta)+r_{2}r_{3}\Big((x_{2}-x_{3})\Phi^{t}_{K^*}(\zeta)
 \nonumber \\&&
+(x_{2}+x_{3})\Phi^{s}_{K^*}(\zeta) \Big)\Big]{\cal
E}_{f}^{1}(t^{1}_{f})+\Big[(x_{2}-r^2_{2})\Phi_{K^*}(\zeta)
 \nonumber \\
&& -r_{2}r_{3}\Big((x_{3}-x_{2})\Phi_{K^*}^{t}(\zeta)
 +(2+x_{2}+x_{3})\Phi_{K^*}^{s}(\zeta)\Big)
\Big] {\cal E}_{f}^{2}(t^{2}_{f})\Big\},
\end{eqnarray}
with $r_{3}=m_{K^*}/m_{B}$ and $\zeta=1-x_{3}$. In our
considerations, the small effects from $r^{2}_{3}\approx 0.028$
and $r^{3}_{2}\approx 0.05$ are neglected. The evolution factors
${\cal E}_{a(f)}^{i}$ are defined by
\begin{eqnarray*}
{\cal
E}^{i}_{a}(t^{i}_{a})&=&\Big(C_{1}(t^{i}_{a})+\frac{C_{2}(t^{i}_{a})}{N_{c}}\Big)\alpha_{s}(t^{i}_{a})
S_{K^*+D_{s}}(t^{i}_{a}) h_{a}(\{x\},\{b\}), \nonumber \\
{\cal
E}^{i}_{f}(t^{i}_{f})&=&\frac{C_{2}(t^{i}_{f})}{N_{c}}\alpha_{s}(t^{i}_{f})
S_{B+D_{s}+K^*}(t^{i}_{f}) h^{i}_{f}(\{x\},\{b\}),
\end{eqnarray*}
where $t^{1,2}_{a,f}$ denote the hard scales and are chosen as
\begin{eqnarray}
t^{1}_{a}&=&
max(m_{B}\sqrt{(1-r^{2}_{2})x_{3}},\frac{1}{b_{2}},\frac{1}{b_{3}}),
\ \ \
t^{2}_{a}=max(m_{B}\sqrt{(1-r^{2}_{2})x_{2}},\frac{1}{b_{2}},\frac{1}{b_{3}}), \nonumber \\
t^{j}_{f}&=&max(m_{B}\sqrt{F^{2}_{j}},m_{B}\sqrt{(1-r^{2}_{2})x_{2}x_{3}},\frac{1}{b_{2}},\frac{1}{b_{3}}),
\nonumber \\
F^{2}_{1}&=&(1-r^{2}_{2})(x_{1}-x_{2})x_{3},\ \ \
F^{2}_{2}=x_{1}+x_{2}+(1-r^{2}_{2})(1-x_{1}-x_{2})x_{3}.
\label{hard-scale}
\end{eqnarray}
Here, $S_{K^*+D_{s}}=S_{K^*}(k^{-}_{3},p^{-}_{3}-k_{3}^{-})
S_{D_{s}}(k^{+}_{2})$ and $S_{B+D_{s}+K^*}=S_{B}(k^{+}_{1})
S_{D_{s}}(k^{+}_{2})S_{K^*}(k^{-}_{3},p^{-}_{3}-k_{3}^{-})$ are
the associated Sudakov factors. We note that only the light
valence quarks of $B$ and $D^{(*)}_{s}$ mesons have the Sudakov
effects \cite{Li}. $h_{a,f}$ describe the hard functions arising from the
propagators of gluon and internal valence quark.
Their detailed expressions with threshold resummation effects can
be found in Ref. \cite{CKL-PRD}.

It is known that besides the longitudinal polarization, there also
involve two transverse polarizations in $B\to VV$ decays
\cite{CKL-PRD2}, where $V$ denotes the vector meson. Therefore, the
$B_{d}\to D_{s}^{*-} K^{*+}$ decay amplitude will be more
complicated than those in $B\to VP$ or $PP$ decays with $P$ being
the pseudoscalar meson. In terms of helicity basis, the
$B_{d}\to D^{*-}_{s} K^{*+}$ decay rate is written as
\begin{equation*}
\Gamma =\frac{G_{F}^{2}P_c}{16\pi m^{2}_{B} } \sum_{h=L,T}{\cal
M}^{(h)\dagger }{\cal M}^{(h)}, \label{dr1}
\end{equation*}
where  the superscript $h$ denotes the helicity states of the two
vector mesons. The amplitude ${\cal M}^{(h)}$ is decomposed into
\begin{eqnarray}
{\cal M}^{(h)} &=&\epsilon_{2h\mu}^{*}\epsilon_{3h\nu}^{*} \left[
a \,\, g^{\mu\nu} + {b \over m_{D^{*}_{s}} m_{K^*}} p_1^\mu
p_1^\nu + i{c \over m_{D^{*}_{s}} m_{K^*}}
\epsilon^{\mu\nu\alpha\beta} p_{2\alpha} p_{3\beta}\right],
\nonumber \\
&\equiv &m_{B}^{2}{\cal M}_{L}+m_{B}^{2}{\cal M}_{N}
\epsilon^{*}_{2T}\cdot\epsilon^{*}_{3T} +i{\cal
M}_{T}\epsilon^{\alpha \beta\gamma \rho}
\epsilon^{*}_{2T\alpha}\epsilon^{*}_{3T\beta} p_{2\gamma }p_{3\rho
}\,,
\end{eqnarray}
with the convention of $tr(\gamma_5\not a\not b\not c\not d)=
-4i\epsilon^{\alpha\beta\gamma\rho}a_\alpha b_\beta c_\gamma
d_\rho$ and the definitions of
\begin{eqnarray}
m_B^2 {\cal M}_L &=& a \,\, \epsilon_{2L}^{*} \cdot
\epsilon_{3L}^{*} +{b \over m_{D^*_{s}} m_{K^*}} \epsilon_{2L}^{*}
\cdot p_1 \, \epsilon_{3L}^{*} \cdot p_1,
\nonumber \\
m_B^2 {\cal M}_N &=& a \,\, \epsilon_{2T}^{*}\cdot
\epsilon_{3T}^{*},
\nonumber \\
{\cal M}_T &=& {c \over m_{D^*_{s}} m_{K^*}}. \label{id-rel}
\end{eqnarray}
Similar to $B_{d}\to D_{s} K^{*+}$, the decay
amplitudes corresponding to each polarizations can be written as
\begin{eqnarray*}
{\cal M}_{L(N,T)}&=&f_{B}F_{L(N,T)} + M_{L(N,T)},
\end{eqnarray*}
where the first (second) term expresses factorized
(nonfactorized) effects. Each hard amplitudes can be
formulated by
\begin{eqnarray}
F_{L} &=&-8\pi C_{F}m_{B}^{2}\int_{0}^{1}d[x]\int_{0}^{\infty} b
d[b]\Phi_{D^*}(x_2)
\nonumber \\
&& \times \Big\{
\left[(r^{2}_{2}+(1-r^{2}_{2})x_{3})\Phi_{K^*}(\zeta)
+r_{2}r_{3}\left(\Phi^{t}_{K^*}(\zeta)-\Phi^{s}_{K^*}(\zeta)\right)\right]
 {\cal E}_{a}^{1}(t^{1}_{a})
\nonumber \\
&& - \left [ (1-r^2_{2})x_{2}\Phi_{K^*}(\zeta)
+2r_{2}r_{3}(1-x_{2})\Phi^{s}_{K^*}(\zeta) \right] {\cal
E}_{a}^{2}(t^{2}_{a}) \Big\},
\end{eqnarray}
\begin{eqnarray}
F_{N} &=&-8\pi C_{F}m_{B}^{2}r_{2}
r_{3}\int_{0}^{1}d[x]\int_{0}^{\infty} b d[b]\Phi_{D^*}^{T}(x_2)
\nonumber \\
&& \times \Big\{ \left[(1+x_{3})\Phi^{v}_{K^*}(\zeta)
+\left(-\frac{r_{2}}{r_{3}}\Phi^{T}_{K^*}(\zeta)+(1-x_{3})\Phi^{a}_{K^*}(\zeta)\right)\right]
 {\cal E}_{a}^{1}(t^{1}_{a})
\nonumber \\
&& - \left [ (1+x_{2})\Phi_{K^*}^{v}(\zeta)
-(1-x_{2})\Phi^{a}_{K^*}(\zeta) \right] {\cal
E}_{a}^{2}(t^{2}_{a}) \Big\},
\end{eqnarray}
\begin{eqnarray}
F_{T} &=&-16\pi C_{F}m_{B}^{2}r_{2}
r_{3}\int_{0}^{1}d[x]\int_{0}^{\infty} b d[b]\Phi_{D^*}^{T}(x_2)
\nonumber \\
&& \times \Big\{ \left[(1-x_{3})\Phi^{v}_{K^*}(\zeta) +\left(
\frac{r_{2}}{r_{3}}
\Phi^{T}_{K^*}(\zeta)+(1+x_{3})\Phi^{a}_{K^*}(\zeta)\right)\right]
 {\cal E}_{a}^{1}(t^{1}_{a})
\nonumber \\
&&- \left [ (1+x_{2})\Phi_{K^*}^{a}(\zeta)
-(1-x_{2})\Phi^{v}_{K^*}(\zeta) \right] {\cal
E}_{a}^{2}(t^{2}_{a}) \Big\},
\end{eqnarray}
\begin{eqnarray}
M_{L} &=&16\pi C_{F}M_{B}^{2}\sqrt{2N_{c}}
\int_{0}^{1}d[x]\int_{0}^{\infty }[b]d[b] \Phi
_{B}(x_{1},b_{1})\Phi_{D^*}(x_{2})
\nonumber \\
&&\times \Big\{ \Big[ (1-2r^{2}_{2})x_3
\Phi_{K^*}(\zeta)+r_{2}r_{3} \Big((x_{3}+x_2)\Phi^{t}_{K^*}(\zeta)
+(x_{2}-x_3 )\Phi^{s}_{K^*}(\zeta) \Big) \Big]{\cal
E}_{f}^{1}(t^{1}_{f})
\nonumber \\
&& + \Big[-(r^{2}_{2}+(1-2r^{2}_{2})x_2)\Phi_{K^*}(\zeta)
+r_{2}r_{3}\Big((2-x_{2}-x_{3})\Phi^{t}_{K^*}(\zeta) \nonumber
\\&& + (x_2-x_3)\Phi^{s}_{K^*}(\zeta) \Big)\Big]{\cal
E}_{f}^{2}(t^{2}_{f}) \Big\},
\end{eqnarray}
\begin{eqnarray}
M_{N} &=&-32\pi C_{F}m_{B}^{2} \sqrt{2N_{c}}r_{2}r_{3}
\int_{0}^{1}d[x]\int_{0}^{\infty }[b]d[b] \Phi _{B}(x_{1},b_{1})
 \Phi_{D^*}(x_2)\Phi^{v}_{K^*}(\zeta)
 {\cal E}_{f}^{2}(t^{2}_{f}),\\
M_{T} &=&-64\pi C_{F}m_{B}^{2} \sqrt{2N_{c}}r_{2}r_{3}
\int_{0}^{1}d[x]\int_{0}^{\infty }[b]d[b] \Phi _{B}(x_{1},b_{1})
 \Phi_{D^*}(x_2)\Phi^{a}_{K^*}(\zeta)
 {\cal E}_{f}^{2}(t^{2}_{f}).
\end{eqnarray}
 From the above equations, we see clearly that the hard amplitudes for
the transverse polarizations are proportional to factors of $r_{2}
r_{3}$ and $r^{2}_{2}$. One of two $r_{2}$ in the latter comes
from the mass of charm-quark $m_{c}$ that we already set $m_{c}
\approx m_{D^{(*)}_{s}}$ due to $m_{D_{s}^{(*)}}-m_{c}\sim
\bar{\Lambda}$.

To get numerical estimations, we model the $B$-meson wave function
$\Phi_{B}$ to be
\begin{eqnarray}
\Phi_{B}(x,b)&=& N_{B}x^{2}(1-x)^{2} \exp\Big[-\frac{1}{2}\Big(
\frac{x
\,m_{B}}{\omega_{B}}\Big)^{2}-\frac{\omega_{B}^{2}b^{2}}{2} \Big]
\end{eqnarray}
where $b$ is the conjugate variable of the transverse momentum of
the light quark, $N_{B}$ can be determined by the normalization of
the wave function at $b=0$ and $\omega_{B}$ is the shape parameter.
Since the wave functions of the $K^*$ meson are derived in the
framework of QCD sum rules, we express them up to twist-3 directly
by \cite{BBKT}
\begin{eqnarray*}
\Phi_{K^* }( x) &=&\frac{3f_{K^*}}{\sqrt{2N_{c}}}x(1-x)
\Big[1+0.57(1-2x)+0.07C^{3/2}_2(1-2x)\Big]\;,
\label{pk2}
\end{eqnarray*}
\begin{eqnarray*}
\Phi_{K^*}^{t}( x) &=&\frac{f_{K^*}^T}{2\sqrt{2N_{c}}} \bigg\{
0.3(1-2x)\left[3(1-2x)^2+10(1-2x)-1\right]+1.68C^{1/2}_4(1-2x)
\nonumber \\
&&+0.06(1-2x)^2\left[5(1-2x)^2-3\right] +0.36\left[
1-2(1-2x)(1+\ln(1-x))\right] \bigg\} \;,
\label{pk3t}
\end{eqnarray*}
\begin{eqnarray*}
\Phi _{K^*}^s( x)  &=&\frac{f_{K^*}^T}{2\sqrt{2N_{c}}} \bigg\{
3(1-2x)\left[1+0.2(1-2x)+0.6(10x^2-10x+1)\right]
\nonumber \\
& &-0.12x(1-x)+0.36[1-6x-2\ln(1-x)]\bigg\} \;,
\label{pk3s}
\end{eqnarray*}
\begin{eqnarray*}
\Phi_{K^*}^{T}(x)&=&\frac{3f_{K^*}^T}{\sqrt{2N_c}} x(1-x)
\Big[1+0.6(1-2x)+0.04C^{3/2}_2(1-2x)\Big]\;,
\label{pkt}\\
\Phi_{K^*}^v(x)&=&\frac{f_{K^*}}{2\sqrt{2N_c}}
\bigg\{\frac{3}{4}\Big[1+(1-2x)^2+0.44(1-2x)^3\Big]
+0.4C^{1/2}_2(1-2x)
\nonumber \\
&& +0.88C^{1/2}_4(1-2x)+0.48[2x+\ln(1-x)] \bigg\}\;,
\label{pkv}\\
\Phi_{K^*}^a(x) &=&\frac{f_{K^*}}{4\sqrt{2N_{c}}}
\Big\{3(1-2x)\Big[1+0.19(1-2x)+0.81(10x^2-10x+1)\Big]\nonumber \\
&&-1.14x(1-x)+0.48[1-6x-2\ln(1-x)]\Big\}\;, \label{pka}
\end{eqnarray*}
with the Gegenbauer polynomials,
\begin{eqnarray*}
C_2^{1/2}(\xi)=\frac{1}{2}(3\xi^2-1)\;,\;\;\;\;
C_4^{1/2}(\xi)=\frac{1}{8}(35 \xi^4 -30 \xi^2 +3)\;,\;\;\;\;
C_2^{3/2}(\xi)=\frac{3}{2}(5\xi^2-1)\;.
\end{eqnarray*}
Although there are theoretical errors on the coefficients of the
wave functions, we find that the allowed errors in these derived
wave functions change the BRs only at a few percent level so that we
will not discuss them further. In the PQCD approach, the wave
functions represent the nonperturbative QCD effects and have the
universal property. In principle, they can be determined by some
specific measured modes. Hence, the unknown $\omega_{B}$ can be
fixed by such as $B\to \phi K $ decays, in which $\phi$ and $K$
meson wave functions have been determined in the literature.
Consequently, the remaining uncertain wave functions are the
$D^{(*)}_{s}$ mesons.

As known that the BR of $B_{d}\to D^{-}_{s} K^{+}$ has been
observed by Belle and Babar so that we can take it as the
criterion to model the relevant $D^{-}_{s}$ wave functions; and
because the mass difference between $D^{-}_{s}$ and $D^{*-}_{s}$
is not much, for simplicity, except the decay constants, their
wave functions are taken to have the same behavior. It is worthful
to mention that as the $b$-dependence on the wave function of $B$
meson, for controlling the size of charmed mesons, we also
introduce the intrinsic $b$-dependence on those of charmed mesons.
Hence, we use the wave functions of $D^{(*)}_{s}$ as
\begin{eqnarray}
\Phi_{D^{(*)}_{s}}(x,b)=\frac{ f_{D^{(*)}_{s}}}{2\sqrt{2N_{c}}}
\Big\{6x(1-x)[1 + c_{D_{s}}\, (1-2x)]\Big\}
\exp\Big[-\frac{\omega_{D_{s}}^{2}b^{2}}{2}\Big],
\end{eqnarray}
where $c_{D_{s}}$ and $\omega_{D_{s}}$ are the unknown parameters.
Although $c_{D_{s}}$ is a free parameter, it can be chosen such
that the $D^{(*)}_{s}$ meson wave functions have the maximum at
$x\approx \bar{\Lambda}/m_{D^{(*)}_{s}}\sim 0.3$ for $m_{c}=1.4$
GeV. And then, we can fix $\omega_{D_{s}}$ through the observation
of the $B_{d}\to D^{-}_{s} K^+$ decay.

In our numerical calculations, the values of theoretical inputs
are set as: $\omega_{B}=0.4$, $f_{B}=0.19$,
$f_{D^{(*)}_{s}}=0.22\, (0.24)$, $f^{(T)}_{K^*}=0.20\, (0.16)$,
$f_{K}=0.16$, $m_{B}=5.28$, $m_{D^{(*)}_{s}}=1.969\, (2.112)$,
$m_{K^*}=0.892$ and $m^{0}_{K}=1.7$ GeV.
With
these  values, we get  $F^{B\to K}=0.35$.
Since the decay $B_{d}\to D^{-}_{s} K^+$ has been studied by Ref.
\cite{LU}, we use the derived formulas in Ref. \cite{LU}
directly. Although the decay $B_{d}\to D^{*-}_{s} K^{+}$ has not
been considered yet, due to the results only related to the
longitudinal part of $D^{*-}_{s}$ which should be similar to the
longitudinal contributions of the $D^{*-}_{s} K^{*+}$ mode, we also
include it in our discussion.
In Table \ref{ha}, we present
the magnitudes of
hard amplitudes.
\begin{table}[htb]
\caption{ The hard amplitudes of $B_{d}\to D^{(*)-}_{s} K^{*+}$
(in units of $10^{-3}$) with $c_{D_{s}}=0.9$ and
$\omega_{D_{s}}=0.45$.} \label{ha}
\begin{ruledtabular}
\begin{tabular}{cccc}
$f_{B}F_{D_{s}^{-} K^{*+}}$ & $M_{D_{s}^{-} K^{*+}}$ &
$f_{B}F_{L}$ & $M_{L}$
\\\hline
  $-0.088-i\,0.125$ & $2.37+i\,8.29$ & $-0.42-i\,0.09$ & $0.51-i\,5.12$ \\\hline
   $f_{B}F_{N}$ & $M_{N}$ & $f_{B}F_{T}$ &
$M_{T}$ \\\hline
  $-0.01+i\,0.15$ & $0.22-i\,0.35$ & $0.23-i\,0.16$ &
  $-0.10-i\,0.24$
\end{tabular}
\end{ruledtabular}
\end{table}
 From the table,
we see that the nonfactorizable
contributions of longitudinal parts are larger than factorizable
ones.
This could be understood by the mechanism of chirality
suppression. The more remarkable example happens in the decays
$B\to \pi \pi$. Because two $\pi$ mesons are identical particles,
under $x_{2}\leftrightarrow x_{3}$ transformation, the decay
amplitude should be the same. However, from the same topology of
Fig. \ref{two-body}a, we know that the four-momentum of the internal
quark in one pion is opposite in sign to that in the another pion.
That means the contributions from hard gluon exchange in opposite
meson sides are canceled each other. Although $D^{(*)-}_{s}$ and
$K^{*+}$ are not identical particles, the cancellation should
still exist. Since the momentum carried by $b$-quark in
nonfactorizable parts, illustrated by Fig. \ref{two-body}b, is
$p_{1}-k_{1}-k_{2}-k_{3}$ and the light quark is
$k_{1}-k_{2}-k_{3}$, the similar cancellations in nonfactorizable
effects are not significant. From the Table \ref{ha}, we also see
that the transverse effects of $B_{d}\to D^{*-}_{s} K^{*+}$
are much smaller than those from longitudinal contributions.

\begin{figure}[hbt]
\includegraphics*[width=2.6
  in]{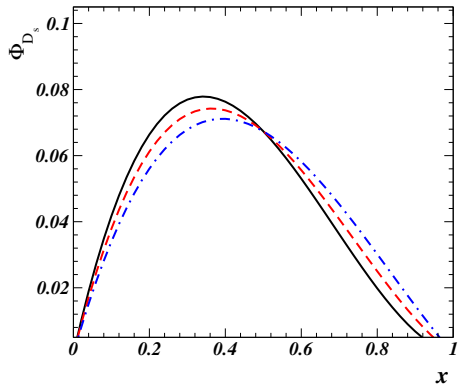} 
\caption{ The wave function of $\Phi_{D_{s}}$ with $c_{D_{s}}=0.9$
(solid line), $c_{D_{s}}=0.7$ (dashed line), and $c_{D_{s}}=0.5$
(dashed-dotted line). }\label{figphids}
\end{figure}

As a result, the predicted BRs of $B_{d} \to D^{(*)-}_{s}
K^{(*)+}$ decays are displayed in Table \ref{cd} (\ref{wd}) for various
values of $c_{D_{s}}$ ($\omega_{D_{s}}$) and $\omega_{D_{s}}=0.45$
($c_{D_{s}}=0.9$). To be more clear, we also show the behavior of
the $D_{s}$ wave function with different values of $c_{D_{s}}$ in
Fig. \ref{figphids}. From the figure, we see that the larger
$c_{D_{s}}$ is, the closer maximum of $\Phi_{D^{(*)}_{s}}$ is to
$x=0.3$.
Since the maximum of the $D_{s}$ wave function is located  around
$0.3$ as shown early, a large $c_{D_{s}}$ is preferred.
Consequently, the BR of $B_{d}\to D_{s}^{-} K^{+}$ intends to be
also large. However, it is necessary to introduce the
$b$-dependence in the $D^{(*)}_{s}$ wave functions  to make the BR
be more adjustable because the accuracy of experimental data is
not good enough. According to our predictions, we find that the
BRs of $B_{d} \to D^{*-}_{s} K^{(*)+}$ decays are smaller than
those of $D^{-}_{s} K^{(*)+}$ modes. Moreover,  if the theoretical
inputs are taken to fit with the observed value of $B_{d}\to
D^{-}_{s} K^{+}$, the BR products of $BR(B_{d}\to D^{(*)}_{s}
K^{*+})\times BR(K^{*+}\to K^{0} \pi^{+})$ should be of ${\cal
O}(10^{-5})$, with $Br(K^{*+}\to K^{0} \pi^{+})=2/3\, Br(K^{*+}\to
(K\pi)^{+})$. Therefore, if the observations of $B_{d}\to
D^{(*)}_{s} K^0 \pi^{+}$ are close to ${\cal O}(10^{-4})$, the
corresponding nonresonant three-body decays will be dominant.
However, if experimental data conclude that the BRs of considered
three-body final states are  larger than our predictions
significantly, but not approaching to $10^{-4}$, it means that the
pure three-body and quasi-two-body decays are comparable. On the
other hand, if our predictions are consistent with the
measurements of experiment, one can conclude that the three-body
decays via the topologies of Fig. \ref{topology} are not important
and negligible.

\begin{table}[htb]
\caption{ The BRs (in units of $10^{-5}$) with the various values
of $c_{D_{s}}$ and $\omega_{D_{s}}=0.45$.} \label{cd}
\begin{ruledtabular}
\begin{tabular}{ccccc}
$c_{D_{s}}$ & $B_{d}\to D^{-}_{s} K^{+}$ & $B_{d}\to D^{*-}_{s}
K^{+}$ & $B_{d}\to D^{-}_{s} K^{*+}$&$B_{d}\to D^{*-}_{s} K^{*+}$
\\\hline
  0.9 & 2.77 & 1.34 & 4.34 & 2.00 \\ \hline
  0.7 & 2.33 & 1.09 & 3.79 & 1.66 \\\hline
  0.5 & 1.94 & 0.87 & 3.32 & 1.39
\end{tabular}
\end{ruledtabular}
\end{table}
\begin{table}[htb]
\caption{ The BRs (in units of $10^{-5}$) with different values of
$\omega_{D_{s}}$ and $c_{D_{s}}=0.9$. }\label{wd}
\begin{ruledtabular}
\begin{tabular}{ccccc}
$\omega_{D_{s}}$ & $B_{d}\to D^{-}_{s} K^{+}$ & $B_{d}\to
D^{*-}_{s} K^{+}$ & $B_{d}\to D^{-}_{s} K^{*+}$&$B_{d}\to
D^{*-}_{s} K^{*+}$
\\\hline
  0.5 & 2.34 & 1.19 & 3.79 & 1.74  \\ \hline
  0.4 & 3.19 & 1.50 & 4.85 & 2.30 \\ \hline
  0.3 & 4.18 & 1.82 & 5.98 & 2.88
\end{tabular}
\end{ruledtabular}
\end{table}

\begin{figure}[hbt]
\includegraphics*[width=1.5
  in]{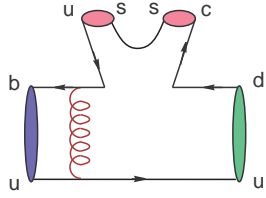} 
\caption{ Topology for nonresonant three-body decays $B^{+}\to
D_{s}^{(*)+} K^{-} \pi^{+}$. }\label{three-body2}
\end{figure}

Finally, some remarks are given. In this paper, we assume that the
mechanism of the color transparency \cite{Bjorken} still dominates in
$B$ decaying to charmed mesons so that we don't have to consider the
rescattering effects. From our results, we know that the PQCD
approach can match the experimental data of the $B_{d}\to D^{-}_{s}
K^+$ decay. With the same approach, our predictions on other
decays, arising also from annihilation topologies, should be
reliable. We note that the relevant charged $B$ decays of $B^{+}\to
D^{(*)+}_{s} \bar{K}^{(*)0}$ are also governed by annihilation
contributions. Due to the suppression of CKM matrix elements,
$V_{ub}V^{*}_{cd}$, the BRs are estimated to be
of ${\cal O}(10^{-8})$ \cite{LU}. Since the corresponding
nonresonant three-body final states $D^{(*)+}_{s} K^{-} \pi^{+}$
are also suppressed by the same CKM matrix elements, illustrated by
Fig. \ref{three-body2}, $B^{+}\to D^{(*)+}_{s} K^{-} \pi^{+}$
decays
cannot give us  more interesting information on three-body decays.
\\

\newpage

\noindent {\bf Acknowledgments:}

The author would like to thank H.N. Li, H.Y. Cheng, C.K. Chua,
K.C. Yang, W.S. Hou, C.D. Lu and K. Ukai for their useful
discussions. This work was supported in part by the National
Science Council of the Republic of China under Grant No.
NSC-91-2112-M-001-053 and the National Center for
Theoretical Sciences of R.O.C.. \\



\begin{references}

\bibitem{Belle-PRD} Belle Collaboration, A. Garmash {\it et al.}, Phys. Rev. D{\bf 65}, 092005
(2002); A.Satpathy {\it et al.}, hep-ex/0211022.

\bibitem{CP}B. Bajc {\it et al.}, Phys. Lett. B{\bf 447}, 313
(1999); A. Deandrea {\it et al.}, Phys. Rev. D{\bf 62}, 036001
(2000); A. Deandrea and A.D. Polosa, Phys. Rev. Lett. {\bf 86},
216 (2001); J. Tandean and S. Gardner, Phys. Rev. D{\bf 66},
034019 (2002).

\bibitem{Chen-PRD}C.H. Chen, Phys. Rev. D{\bf 67},014012 (2003).

\bibitem{CHT} C.K. Chua, W.S. Hou and S.Y. Tsai, Phys. Lett. B{\bf
544}, 139 (2002).


\bibitem{ChenLi} C.H. Chen and H.N. Li, hep-ph/0209043.

\bibitem{CY} H.Y. Cheng and K.C. Yang, Phys. Rev. D{\bf 66},
054015 (2002).

\bibitem{Belle-PLB} A. Drutskoy {\it et al.}, Phys. Lett. {\bf B} 542, 171
(2002).


\bibitem{Balakin}
V.E. Balakin {\it et al.}, Phys. Lett. B{\bf 41}, 205 (1972); M.
Bernardini {\it et al.}, Phys. Lett. B{\bf 44}, 393 (1973); {\bf
46}, 261 (1973); B. Delcourt {\it et al.}, Phys. Lett. B{\bf 99},
257 (1981); P.M. Ivanov {\it et al.}, Phys. Lett. B {\bf 107}, 297
(1981); DM2 Collaboration, D. Bisello {\it et al.}, Z. Phys. C{\bf
39}, 13 (1988); F.~Mane {\it et al.}, Phys. Lett. B{\bf 99}, 261
(1981); S.I. Dolinsky {\it et al.}, Phys. Rept. {\bf 202}, 99
(1991); R.~R.~Akhmetshin {\it et al.}, Phys. Lett. B{\bf 364}, 199
(1995).

\bibitem{CHST}C.K. Chua, W.S. Hou, S.Y. Shiau and S.Y. Tsai,
hep-ph/0209164.

\bibitem{MP} D. Muller et al., Fortschr. Physik. {\bf 42}, 101 (1994);
M. Diehl, T. Gousset, B. Pire, and O. Teryaev, Phys. Rev. Lett.
{\bf 81}, 1782 (1998); M.V. Polyakov, Nucl. Phys. {\bf B555}, 231
(1999).

\bibitem{Belle} Belle Collaboration, P. Krokovny {\it et al.},
Phys. Rev. Lett. {\bf 89}, 231804 (2002).

\bibitem{Babar} Babar Collaboration, B. Aubert, {\it et al.},
hep-ex/0207053.

\bibitem{CKM}  N. Cabibbo, Phys. Rev. Lett. {\bf 10}, 531 (1963); M.
 Kobayashi and T. Maskawa, Prog. Theor. Phys. {\bf 49}, 652 (1973).

\bibitem{BBL}  G. Buchalla , A.J. Buras and M.E. Lautenbacher,
Rev. Mod. Phys. {\bf 68}, 1230 (1996).

\bibitem{LB} G.P. Lepage and S.J. Brodsky, Phys. Lett. B{\bf 87}, 359
(1979); Phys. Rev. D{\bf 22}, 2157 (1980).

\bibitem{Li} H.N. Li, Phys. Rev. D{\bf 64}, 014019 (2001).

\bibitem{KLS-PRD} T. Kurimoto, H.N. Li and A.I. Sanda, Phys. Rev. D{\bf
65}, 014007 (2002).

\bibitem{CKL-PRD} C.H. Chen, Y.Y. Keum and H.N. Li, Phys. Rev.
D{\bf 64}, 112002 (2001).

\bibitem{KLS}  Y.Y. Keum, H.N. Li, and A.I. Sanda, Phys. Lett. B{\bf 504},
6 (2001); Phys. Rev. D{\bf 63}, 054008 (2001).

\bibitem{LUY}  C.D. L${\rm \ddot{u}}$, K. Ukai, and M.Z. Yang, Phys. Rev. D%
{\bf 63}, 074009 (2001).

\bibitem{CL-PRD} C.H. Chen and H.N. Li, Phys. Rev. D{\bf 63}, 014003 (2001).

\bibitem{Chen-PLB520}C.H. Chen, Phys. Lett. B{\bf 520}, 33 (2001).

\bibitem{Melic} B. Melic, Phys. Rev. D{\bf 59}, 074005 (1999).

\bibitem{KS} E. Kou and A.I. Sanda, Phys. Lett. B{\bf 525}, 240 (2002).

\bibitem{Chen-PLB525} C.H. Chen, Phys. Lett. B{\bf 525}, 56
(2002).

\bibitem{CQ-PRD} C.H. Chen and C.Q. Geng, Phys. Rev. D{\bf 66},
094018 (2002).

\bibitem{SU} A.I. Sanda and K. Ukai, Prog. Theor. Phys. {\bf 107}, 421 (2002).

\bibitem{Keum} Y.Y. Keum, H-n. Li, and A.I. Sanda, AIP Conf.Proc. 618 (2002), 229;
Y.Y. Keum, hep-ph/0209002; Y.Y. Keum and A.I. Sanda,
hep-ph/0209014.

\bibitem{Li-P}H.N. Li, presented at FPCP, Philadelphia, Pennsylvania, 16-18 May 2002,
hep-ph/0210198; T. Kurimoto, H.N. Li and A.I. Sanda,
hep-ph/0210289.

\bibitem{BBKT} P. Ball, V.M. Braun, Y. Koike, and K. Tanaka, Nucl. Phys. B{\bf 529}, 323 (1998).

\bibitem{CKL-PRD2} C.H. Chen, Y.Y. Keum and H.N. Li, Phys. Rev. D{\bf 66}, 054013 (2002).

\bibitem{LU} C.D. L${\rm \ddot{u}}$ and K. Ukai, hep-ph/0210206.

\bibitem{Bjorken} J.D. Bjorken, Topics in $B$-physics, Nucl. Phys.
{\bf 11} ({\it Proc. Suppl.}), 325 (1989).

\end{references}
\end{document}